\documentclass{aa}  

\usepackage{graphicx}
\usepackage{txfonts}
\usepackage{natbib}
\usepackage{multirow}
\usepackage{hyperref}
\usepackage{adjustbox}
\usepackage{subcaption}
\usepackage{subcaption}
\usepackage{amsmath}
\usepackage{ulem}
\usepackage[usenames,dvipsnames,table]{xcolor}
\hypersetup{
    colorlinks=true,
    linkcolor=blue,
    filecolor=blue,      
    urlcolor=blue,
    citecolor=blue
    }

\def\be{\begin{equation}}
\def\ee{\end{equation}}
\def\ba{\begin{eqnarray}}
\def\ea{\end{eqnarray}}

\graphicspath{ {Figures/} }
\bibpunct{(}{)}{;}{a}{}{,} 
\begin{document} 

   \title{Distinguishing coupled dark energy models with neural networks}

   \author{L. W. K. Goh
          \inst{1}\fnmsep\thanks{Both authors contributed equally}
          \and
          I. Ocampo\inst{2}\fnmsep\footnotemark[1]
          \and
          S. Nesseris\inst{2}
          \and 
          V. Pettorino\inst{3}
          }

   \institute{Université Paris-Saclay, Université Paris Cité, CEA, CNRS, Astrophysique, Instrumentation et Modélisation Paris-Saclay, 91191 Gif-sur-Yvette, France\\
              \email{lisa.goh@cea.fr}
         \and
             Instituto de Física Téorica UAM-CSIC, Universidad Autónoma de Madrid, Cantoblanco, 28049 Madrid, Spain
        \and
             European Space Agency/ESTEC, Keplerlaan 1, 2201 AZ Noordwijk, The Netherlands
             }

   \date{Received XXXX; accepted YYYY}

 
  \abstract
   {}
   {
   We investigate whether neural networks (NNs) can accurately differentiate between growth-rate data of the large-scale structure (LSS) of the Universe simulated via two models: a cosmological constant and $\Lambda$ cold dark matter (CDM) model and a tomographic coupled dark energy (CDE) model.}
   {We built an NN classifier and tested its accuracy in distinguishing between cosmological models. For our dataset, we generated $f\sigma_8(z)$ growth-rate observables that simulate a realistic Stage IV galaxy survey-like setup for both $\Lambda$CDM and a tomographic CDE model for various values of the model parameters. We then optimised and trained our NN  with \texttt{Optuna},
   aiming to avoid overfitting and to maximise the accuracy of the trained model. We conducted our analysis for both a binary classification, comparing between $\Lambda$CDM and a CDE model where only one tomographic coupling bin is activated, and a multi-class classification scenario where all the models are combined.}
   {For the case of binary classification, we find that our NN can confidently (with $>86\%$ accuracy) detect non-zero values of the tomographic coupling regardless of the redshift range at which coupling is activated and, at a $100\%$ confidence level, detect the $\Lambda$CDM model. For the multi-class classification task, we find that the NN performs adequately well at distinguishing $\Lambda$CDM, a CDE model with low-redshift coupling, and a model with high-redshift coupling, with 99\%, 79\%, and 84\% accuracy, respectively.}
  {By leveraging the power of machine learning, our pipeline can be a useful tool for analysing growth-rate data and maximising the potential of current surveys to probe for deviations from general relativity.}


   \maketitle
%

\section{Introduction}
In the past few decades, advances in observational cosmology have allowed us to achieve percent-level precision, bringing us ever closer to understanding the nature of the  Universe. Currently, the concordance model postulates the presence of ordinary baryonic matter, slow-moving and non-interacting (hence cold) dark matter (DM), and a cosmological constant, $\Lambda$, assumed to be responsible for the late-time accelerating expansion. Collectively, this is known as the $\Lambda$ cold dark matter (CDM) model and has been largely successful in providing an accurate description of the Universe. However, with the recent continuous releases of observational data, tensions have arisen \citep{DIVALENTINO2021102605}, most notably the significant discrepancy (of an order of $\sim5\sigma$) between the measured, current-day value of the expansion of the Universe -- the Hubble constant ($H_0$) -- derived from high-redshift and low-redshift probes. Putting aside systematic errors in the measurements, this could hint at new physics beyond $\Lambda$CDM that is yet to be uncovered. 

As such, a variety of alternative cosmological models have been proposed \citep{Schoneberg:2021qvd} in an attempt to answer the questions that $\Lambda$CDM has so far been unable to. This typically involves modifying Einstein's field equations, either by generalising the Einstein-Hilbert Lagrangian to modify the space-time geometry \citep{CLIFTON20121} or by altering the matter-energy contents of the Universe. In this study, we focused on a class of models that deal with the latter. Specifically, we assumed that the dark energy component of the Universe exists in the form of a scalar field, mediating a fifth force attraction between the CDM particles \citep{Wetterich:1994bg, Amendola_2000}. The coupled dark energy (CDE) model, as it is known, has been demonstrated to relieve the $H_0$ tension \citep{PhysRevD.101.123513} while being compatible with current data \citep{Pettorino:2013oxa,DiValentino:2019ffd}, making it one of the $\Lambda$CDM model extensions still being actively investigated today \citep{Gomez-Valent:2022bku,Goh_2023}. However, there are also many other variations of CDE models, mainly with different couplings (see for example \citealt{Gumjudpai:2005ry,Gavela:2010tm,Salvatelli:2013wra,Amendola:2006dg,Pourtsidou:2013nha}, and references therein).

To test the CDE model, we availed ourselves of the information on the late-time evolution of the Universe, which is embedded in combined probes of its large-scale structure (LSS). 

Through measurements of galaxy clustering and weak lensing, LSS surveys such as the Kilo Degree Survey \citep[KiDS;][]{de2013kilo}, the Dark Energy Survey \citep[DES;][]{abbott2005dark}, the Dark Energy Spectroscopic Instrument \citep[DESI;][]{levi2019dark}, and others have provided precision in cosmological parameter estimation of better than 5$\%$. 

Measurements at the sub-percent level are expected to be achieved with the current generation of LSS observational programmes, such as the European Space Agency's \textit{Euclid} mission \citep{2011arXiv1110.3193L}  and the Legacy Survey of Space and Time \citep[LSST;][]{2019ApJ...873..111I}. This implies that the upcoming decade is poised to witness an exponential increase in the quantity, variety, and quality of multi-wavelength astronomical observations of the LSS, which will require very sophisticated computational resources. Likewise, at this stage, the constraints may come from the statistical and data-driven tools themselves, rather than the data quality or quantity. As a result, machine learning (ML) techniques have proven to be a valuable tool capable of addressing some of the computational limitations of conventional statistical methods \citep{moriwaki2023machine, kacprzak2022deeplss}.

In this work, our goal is to test slight deviations in the growth history of the Universe with respect to $\Lambda$CDM within the framework of the CDE model. For this purpose, we analysed growth-rate ($f\sigma_8$) data by leveraging ML techniques to 
differentiate between the models. Similar analyses have been carried out for $\Lambda$CDM and other more generalised modified gravity models \citep{Peel:2018aei,Merten:2018bgr,Mancarella2020jyu,Thummel2024nhv,Murakami:2023qdl}.\ We aim to build upon these efforts by investigating how accurately deep learning methods can detect hints of beyond-$\Lambda$CDM physics. 

This paper is arranged as follows: In Sect. \ref{sect:cde} we explain the theoretical framework of the CDE model, specifically a tomographic CDE model, highlighting how the presence of coupling between DM and dark energy (DE) modifies the cosmological observables. We then describe the development of our neural network (NN) architecture as well as the generation of mock data in Sect. \ref{sect:nn}. We present and discuss our results in Sect. \ref{sec:results}, for both a binary and a multi-class classification scenario, and finally summarise our work in Sect. \ref{sect:conclusion}.
\section{Coupled dark energy}\label{sect:cde}
In CDE cosmologies, a scalar field, $\phi,$ is assumed to play the role of DE by driving the late-time accelerating expansion of the Universe. This field consequently mediates an interaction between Fermionic DM particles, resulting in them experiencing a fifth force that can be stronger than gravity. 

In the case of cosmologies with a non-minimal coupling between a scalar field and non-relativistic particles, the general form of the Lagrangian can be split into several components \citep{Koivisto:2015qua}:
\begin{equation}
    \mathcal{L} = \mathcal{L}_{\rm grav} + \mathcal{L}_{\phi} + \mathcal{L}_{\rm dm} 
,\end{equation}
where $\mathcal{L}_{\rm grav}$ is the Einstein-Hilbert Lagrangian, $\mathcal{L}_{\rm dm}$ is that of the DM component, and $\mathcal{L}_{\phi}$ is that of the scalar field given by $\mathcal{L}_{\rm \phi}=\frac{1}{2}\partial_\mu\phi\,\partial^\mu\phi-V(\phi)$, where $V(\phi)$ is the potential of the scalar field.

We can further break down the DM Lagrangian term by expressing it as a sum of its kinetic and interaction terms:
\begin{equation}
    \mathcal{L}_{\rm dm} = \mathcal{L}_{\rm kin} + \mathcal{L}_{\rm int} = \bar{\psi}i\gamma^\mu\partial_\mu\psi + m_{\rm dm}(\phi)\psi\bar{\psi}
,\end{equation}
where $\psi$ refers to the DM wave vector and $m_{\rm dm}$ its field-dependent mass term. Hence, we see that in the case of coupled cosmologies that we are studying, the DE-DM interaction is mediated by the mass of the DM particles. Furthermore, we can specify the form of the mass term as $m(\phi)=m_0e^{\beta\kappa\phi}$, where  $\kappa=\sqrt{8\pi G}$ is the reduced Planck mass, and $\beta\equiv-\frac{1}{\kappa}\frac{\partial \ln{m_{\rm dm}}}{\partial \phi}$ is the coupling parameter quantifying the strength of coupling between the $\phi$ and DM sectors.

From the Lagrangian, we can also derive the energy-momentum tensors $T_{\mu\nu}\equiv\frac{-2}{\sqrt{-g}}\frac{\delta(\sqrt{-g}\mathcal{L})}{\delta g^{\mu\nu}}$ and subsequently construct the modified covariant equations for the $\phi$ and DM components:
\begin{equation}\label{eq:covCons}
    \nabla^\mu T_{\mu\nu}^{\phi}=\kappa\beta T^{\rm dm}\nabla_\nu\phi\quad ;\quad  \nabla^\mu T_{\mu\nu}^{\rm dm}=-\kappa\beta T^{\rm dm}\nabla_\nu\phi\,, 
\end{equation}
with $T^{\rm dm}=g^{\mu\nu}T_{\mu\nu}^{\rm dm}$. For a comprehensive review of the theoretical formalism of the CDE model, we refer the reader to \citet{Amendola_2000} and \citet{Pettorino_2008}.

\subsection{Tomographic coupled dark energy}
A tomographic CDE model, first introduced in \cite{Goh_2023}, is an extension of the CDE model that allows for a coupling that varies with redshift, $z$. For simplicity, we employed in this work a three-bin parameterisation of the coupling strength, where $\beta(z)$ is given by
\begin{equation}\label{eq:binbeta}
\beta(z) = \frac{\beta_1+\beta_n}{2}+\frac{1}{2}\sum_{i=1}^{n-1}(\beta_{i+1}-\beta_{i})\tanh[s(z-z_i)]\,,
\end{equation}
with $n=3$ and redshift bin edges $z_i=\{0,100,1000\}$. To ensure a smooth transition of the coupling between redshift bins, we defined a smoothing factor $s = 0.03$, which is the same value used in \citet{Goh_2023}. In the following equations, we drop the explicit $z$ dependence of $\beta$ for clarity of notation and assume that $\beta_i$ refers to the coupling parameter in the $i$th bin. 

Assuming a flat Friedmann-Lemaître-Robertson-Walker metric where DM and the scalar field behave as perfect fluids, we can solve the Einstein field equations with a modified $T^{\rm dm;\phi}_{\mu\nu}$, arriving at the relevant conservation equations for DM and $\phi$:
\begin{equation}\label{eq:consrhoDM_cde}
\rho^\prime_{\rm dm}+3\mathcal{H}\rho_{\rm dm}=-\kappa\beta\rho_{\rm dm}\phi^{\prime}\,,
\end{equation}

\begin{equation}\label{eq:consrhoscf_cde}
\rho^\prime_{\phi}+3\mathcal{H}(\rho_{\phi}+p_\phi)=\kappa\beta\rho_{\rm dm}\phi^{\prime}\,,
\end{equation}
where the prime denotes a derivative with respect to conformal time $\tau$ and $\mathcal{H}\equiv a'/a$. 

We can also derive the energy density and pressure terms of the scalar field, $\rho_\phi$ and $p_\phi$:
\begin{equation}\label{eq:scf_rho_p}
\rho_\phi=\frac{(\phi^\prime)^2}{2a^2}+V(\phi)\quad ;\quad p_\phi=\frac{(\phi^\prime)^2}{2a^2}-V(\phi)\,,
\end{equation}
and by substituting Eq. \eqref{eq:consrhoscf_cde} into Eq. \eqref{eq:scf_rho_p}, we finally obtain the modified Klein-Gordon equation governing the evolution of the scalar field: 
\begin{equation}\label{eq:KG}
\phi^{\prime\prime}+2\mathcal{H}\phi^{\prime}+a^2\frac{\partial V}{\partial\phi}=\kappa \beta a^2\rho_{\rm dm}\,.
\end{equation}
For simplicity, we adopted a flat potential for the scalar field, where $V(\phi)=V_0$,
that is responsible for late-time acceleration.

\subsection{Perturbation equations}
In a tomographic CDE model, the evolution of the DM density contrast $\delta_{\rm dm}\equiv\delta\rho_{\rm dm}/\rho_{\rm dm}$, can be approximated, at sub-horizon scales, as a second-order ordinary differential equation:

\begin{equation}\label{eq:densityContrastDM}                  \begin{aligned}
\delta_{\rm dm}^{\prime\prime}+&\left[\mathcal{H}-\beta\kappa\phi^\prime\right] \delta^\prime_{\rm dm}
 -\frac{\kappa^2a^2}{2}\left[\rho_{\rm b}\delta_{\rm b}+\rho_{\rm dm}\delta_{\rm dm}(1+2\beta^2)\right]=0\,,
\end{aligned}
\end{equation}
where the subscript `b' denotes baryons.

\begin{figure}[t!]
\centering
\includegraphics[width=\linewidth]{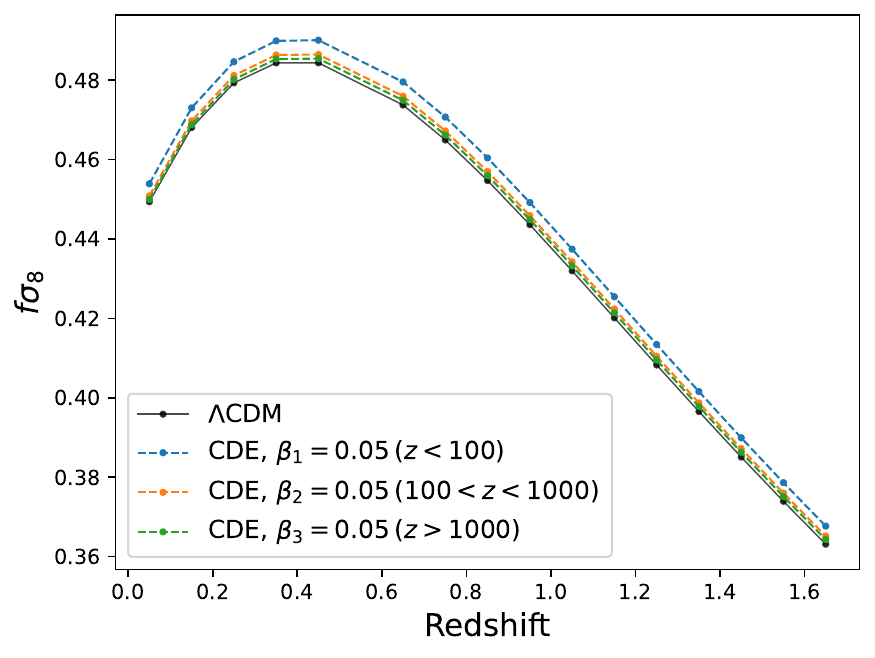}
\caption{Growth rate ($f\sigma_8$) against redshift ($z$) for the 16 redshift bins for both the $\Lambda$CDM model (solid black line) and the CDE models (coloured lines). We used the same cosmological parameters in all models, but in the CDE case set the coupling at the three different redshift bins $z=[0,100,1000]$ to $\beta_i=0.05$.}
\label{fig:fsigma8}
\end{figure}

The growth rate $f$, defined by $f\equiv\frac{d \ln{\delta_{\rm dm}}}{d \ln{a}}$, is the derivative of the logarithm of matter overdensity with respect to the logarithm of the scale factor, which we took to be scale-independent in this model. Since the evolution of $\delta_{\rm dm}$ is modified with respect to $\Lambda$CDM, we expect that this will change the value of $f$ such that $f_\mathrm{eff}=f-\kappa\phi'\beta/\mathcal{H}$. However, we verified that the second contribution term is of the order of $10^{-3}f$, which is within the order of magnitude of the uncertainty of $f\sigma_8$. By solving Eq. \eqref{eq:densityContrastDM}, we can see how coupling modifies clustering dynamics: $\delta_{\rm dm}$ increases due to the presence of a $\beta$ term, and hence we would expect an increase in $f$ and $\sigma_8(z)$, the rms amplitude of clustering at 8$h^{-1}$Mpc. This is further illustrated in Fig. \ref{fig:fsigma8}, where we also see that activating coupling at late times ($z < 100$) has the largest impact on the increase in clustering.

We employed the observable $f\sigma_8(z)$, the product of the growth rate $f$ and $\sigma_8(z)$, as our data. Current Stage IV spectroscopic galaxy surveys such as DESI and \textit{Euclid} will be able to provide values of $f\sigma_8(z)$ through measurements of redshift-space distortions, with forecasts of $\sigma(f\sigma_8)/f\sigma_8$ at less than 5\% \citep{DESI:2016fyo}. It could hence be an effective probe in testing for deviations in the growth history of the Universe with respect to $\Lambda$CDM.   

\section{Numerical analysis}\label{sect:nn}
\subsection{Generating mock DESI-like data}
We used the Boltzmann code \texttt{CLASS} \citep{Diego_Blas_2011} to generate 4000 training and 750 testing datasets of the product $f\sigma_8(z)$ for each model, varying $\Omega_m$ within the range $\Omega_m=[0.01,0.7]$ and fixing the fiducial cosmology to $\omega_b=0.02225$, $\ln10^{10}A_s = 3.044$, $n_s = 0.966$, $\tau= 0.0522$, $V_0 = 2.64\cdot10^{-47}\text{GeV}^4$, where in $\Lambda$CDM, $V_0=\rho_\Lambda$. To be conservative, we set $k_\mathrm{max}=0.1\,h/$Mpc to exclude highly non-linear scales and only considered the linear matter power spectrum.

For the tomographic CDE model, we employed a modified version of \texttt{CLASS}\footnote{\url{https://github.com/LisaGoh/CDE}} to generate the same number of training and testing datasets, varying $\Omega_m$ and additionally, $\beta_i=[0.001,0.5]$\footnote{Note that the value of $\beta$ is rescaled by a factor of $\sqrt{2}$ in the code. We also note that in the case of the modified version of \texttt{CLASS}, $\Omega_m$ refers to the value of the initial matter density, at $z_{\rm ini}=10^{14}$. } for each tomographic bin, separately.  Then we simulated a DESI-like setup of 16 redshift bins with mean redshifts equally spaced between $z=[0.05,1.65]$ and used values of the galaxy bias $b(z)$ and galaxy number densities $dn/dz$ as specified in \cite{DESI:2016fyo}. 

We also included, in our training and testing data, uncertainties in the $f\sigma_8(z)$ measurement within each redshift bin. To do this, we built a covariance matrix by first calculating the Fisher matrix for the observed galaxy power spectrum in each redshift bin $z_i$ \citep{yahia2021validating, tegmark1997measuring}, given by
\ba
F_{\alpha\beta}(z_i)&=&\frac{1}{8\pi^2}\int_{-1}^{1}d\mu\int_{k_{min}}^{k_{max}}k^2dk\left[\frac{\partial P_{\delta\delta}(k,\mu;z_i)}{\partial\alpha}\frac{\partial P_{\delta\delta}(k,\mu;z_i)}{\partial\beta}\right]\nonumber\\ &\times&V_{\rm eff}(z_i;k,\mu),
\ea
where $\alpha$ and $\beta$ are the parameters of concern, $P_{\delta\delta}$ is the linear matter power spectrum and $V_{\rm eff}$ is the effective volume of the survey. We calculated the power spectrum using \texttt{CLASS}, then evaluated its derivatives via a two-point central difference formula with respect to the cosmological parameters, $\theta_{\text{cosmo}}=\left\{\Omega_{\mathrm{b}, 0}, \Omega_{\mathrm{m}, 0}, h, n_s, \ln \left(10^{10} A_s\right), \sum m_\nu\right\}$, and the nuisance parameters, $\theta_{\text{nuis}}=\left\{\sigma_p, P_s\right\}$ (related to the non-linear component of the power spectra and the shot noise). Subsequently, we obtained the total Fisher matrix summing over all the redshift bins as
\begin{equation}
F_{\alpha \beta}=\sum_{i=1}^{N_{\mathrm{bins}}} F_{\alpha \beta}^{\mathrm{bin}}\left(z_i\right)
.\end{equation}

Finally, the Fisher matrix is projected from the $\theta_{\text{cosmo}}$ and $\theta_{\text{nuis}}$ parameters to $f\sigma_8(z_i)$ in the redshift bins and then inverted to obtain the covariance matrix $\textbf{C}_{f\sigma_8f\sigma_8}$, which is used to generate an additional Gaussian sampled noise component added to the values of $f\sigma_8(z)$ output by the Boltzmann code.
This method was compared with other approaches, for example by \cite{blanchard2020euclid} and  \cite{ yahia2021validating}, who find excellent agreement, providing confidence in its robustness.

\subsection{Neural network architecture \label{sec:NN_arc}}
Neural networks are a popular ML technique that simulates the learning mechanism of biological systems, by extracting information from relationships and patterns from data. Every neuron (unit) has a weighted connection with another one, and the architecture depends on the problem 
at hand but typically consists of hidden layers of neurons between the input and output ones. Some packages have been developed to optimise, test, and choose the most appropriate architecture; for the present study, we worked with \texttt{Optuna}\footnote{\url{https://github.com/optuna/optuna}} \citep{optuna} to find the number of layers, the number of neurons in each layer and the values of some hyperparameters. This agnostic framework works by training the same dataset several times with different architectures and extracting the one with the highest accuracy and lowest loss.

\begin{figure}[t!]
    \centering
    \includegraphics[width=0.55\linewidth]{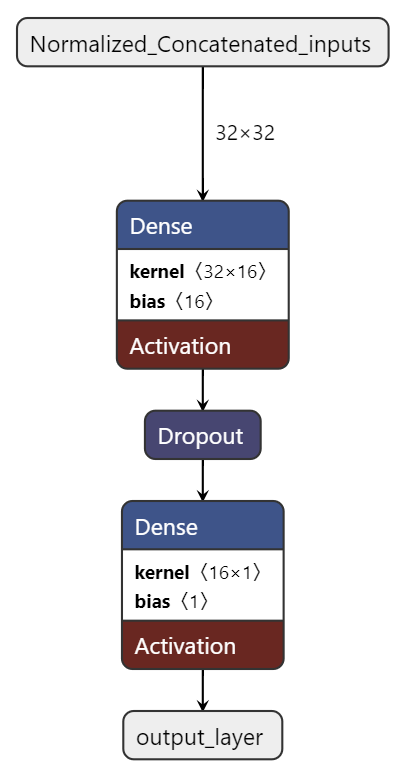}
    \caption{Schematic of the NN architecture we implemented in our work. The normalisation of features and their concatenation as an input array was performed within the architecture. 32$\times$32, as denoted beside the topmost arrow, represents the 32 features (16 data points of $f\sigma_8(z)$ with its standard deviation), with a batch size of 32.}
    \label{fig:NN_architecture}
\end{figure}

Since the goal is to test deviations from $\Lambda$CDM using the NN, we worked with three different architectures designed to discriminate growth-rate  ($f\sigma_8$) data coming from the two models being considered at hand: $\{\Lambda$CDM, CDE($\beta_1)\}$, $\{\Lambda$CDM, CDE($\beta_2)\},$ and $\{\Lambda$CDM, CDE($\beta_3)\}$; we note that we varied each tomographic bin independently of the others. The optimal number of hidden layers for the three cases was 1, while the number of units and the dropout rate varied for each case. We also implemented an early stopping callback with a patience setting of 50 epochs to find the ideal number of epochs to train each architecture and prevent overfitting. This information is available in Table~\ref{tab:hyperparams}, and the generalised diagram of the architecture used is displayed in Fig.~\ref{fig:NN_architecture}, which was generated with \texttt{Netron}\footnote{\url{https://github.com/lutzroeder/netron}}.

On the other hand, 32 is the number of nodes in the input layer, which accounts for the 16 $f\sigma_8(z)$ values and its standard deviation $\sigma(f\sigma_8(z)$), corresponding to each $z$ bin. We implemented feature normalisation with a batch size of 32 \citep{singh2022feature}. After the input layer, we have the hidden layer with $n$ number of units, which varies for each $\beta_i$ parameter (found by \texttt{Optuna}). The activation function of this hidden layer is the Rectified Linear Unit (ReLU; \citep{he2018relu}), 
followed by a dropout layer (commonly used in the literature as a regularisation technique; \citealt{srivastava2014dropout}). Lastly comes the output layer, with a sigmoid activation function to enable the classification task: `0' for $\Lambda$CDM and `1' for CDE. We compiled the models using an Adam optimiser and a binary cross-entropy loss function, while the NN was built using \texttt{TensorFlow Keras} \citep{chollet2015keras}.

Finally, we performed robustness tests on the NN performance. First, we verified the impact of randomness during the training procedure: we trained and tested the network with the same architecture and datasets multiple times, and determined the performance error. We find a performance of $94.4\% \pm 0.2\%$ for the $\beta_{1}$ architecture, $94.1\% \pm 0.1\%$ in the $\beta_{2}$ case and $93.9 \% \pm 0.2\%$ for $\beta_{3}$. This illustrates a high overall level of accuracy in distinguishing between the two models used in our analysis, with a very small error across all architectures. Second, we tested the randomness of dividing the training and test sets with different random seeds, then training and testing the NNs, and found a similar error of $\sim 0.2\%$. These tests confirm that our NN architecture is robust to randomness. 

Finally, we investigated the impact of increasing the number of training datasets. This analysis is relevant for the scalability tests of our NN, since in cosmology the datasets are becoming increasingly large and complex due to advancements in observational technology and simulations. The time complexity of the NN scales approximately linearly with the size of the datasets  (the number of simulated training data), meaning that larger datasets result in proportionally longer training and testing. However, a larger dataset does not scale proportionally with the accuracy or performance of the NN, because the accuracy reaches a saturation point with respect to the size of the dataset. In other words, larger datasets lead to longer training times without proportional improvements in performance. We find 8000 to be the optimal number of training dataset samples, (with a fairly reasonable training and testing time of 60 minutes). For more details, we refer the reader to Appendix \ref{app:a}. A further important consideration is that as the dataset size increases significantly, there is a risk of encountering degeneracies between $\Lambda$CDM and CDE $f\sigma_8$ data, which could negatively impact the NN’s performance and may require more complex architectures. These degeneracies could depend on how we choose to sample our parameter space when generating the data (for example if we decrease the couplings $\beta_1$, $\beta_2$, and $\beta_3$ enough to be very close to $\Lambda$CDM). At the same time, a dataset that is too small may fail to capture a representative range of scenarios, limiting the NN’s ability to learn effectively.

In terms of testing data complexity, when real data become available from current Stage-IV surveys, we do not envision the number of redshift bins to differ greatly from the setup we have adopted. However, we expect a much more realistic and complex covariance matrix compared to the one we employed, which was based on a simplistic Fisher matrix forecast. Hence, this could impact the $f\sigma_8$ uncertainty values and in turn, the ability of the NN to capture the more complex correlations between redshift bins.

\section{Results and discussion\label{sec:results}}
\subsection{Separate $\beta$\label{sec:results_beta}}
In this section, we present the results of our NN for the case where we switched on coupling in only one of the three tomographic bins, $\beta_1$, $\beta_2$, or $\beta_3$, which correspond to the tomographic bins $z<100$, $100<z<1000$, and $z>1000,$ respectively. Hereafter, for brevity, we refer to the model where the coupling is activated within the $i-$th tomographic bin as the $\beta_i$ model.

We first present in Table \ref{tab:hyperparams}, the best-fit hyperparameters obtained by \texttt{Optuna}  and the final number of training epochs when early stopping is invoked. We see that in all cases, having one hidden layer is sufficient, although the number of nodes within the layer can vary widely between models. Interestingly, the best-fit dropout rate is relatively low, at around $20\%$ for all cases. To verify that \texttt{Optuna} indeed gives a better result, we also conducted the same training and testing with an un-optimised architecture, which we describe in more detail in Appendix \ref{app:b}.

\begin{table}[h]
\renewcommand{\arraystretch}{1.2}
  \centering
 \caption{Best-fit hyperparameters as obtained by \texttt{Optuna}.}
  \begin{tabular}{| c | c | c |  c  | c | }

    \hline \hline
      \multirow{3}{*}{Model} & \multicolumn{4}{c|}{Hyperparameter} \\
      \cline{2-5}
                             &Hidden & Nodes & Dropout & Training\\
                            &layers& $(n)$ & rate & epochs\\
     \hline
    $\beta_1$  &1 & 38& 0.224 & 660 \\
    \hline
    $\beta_2$  &1& 116 &0.218 & 683 \\
    \hline
    $\beta_3$  &1& 82 &0.215 & 673\\
    \hline \hline
  \end{tabular}
\tablefoot{Number of hidden layers, number of nodes in the hidden layer, drop-out fraction, as well as the number of training epochs when utilising early stopping, for each tomographic bin that was activated.}
  \label{tab:hyperparams}
\end{table}

\begin{figure}[t!]
\centering
\includegraphics[trim={0 5cm 0 2cm},clip,width=\linewidth]{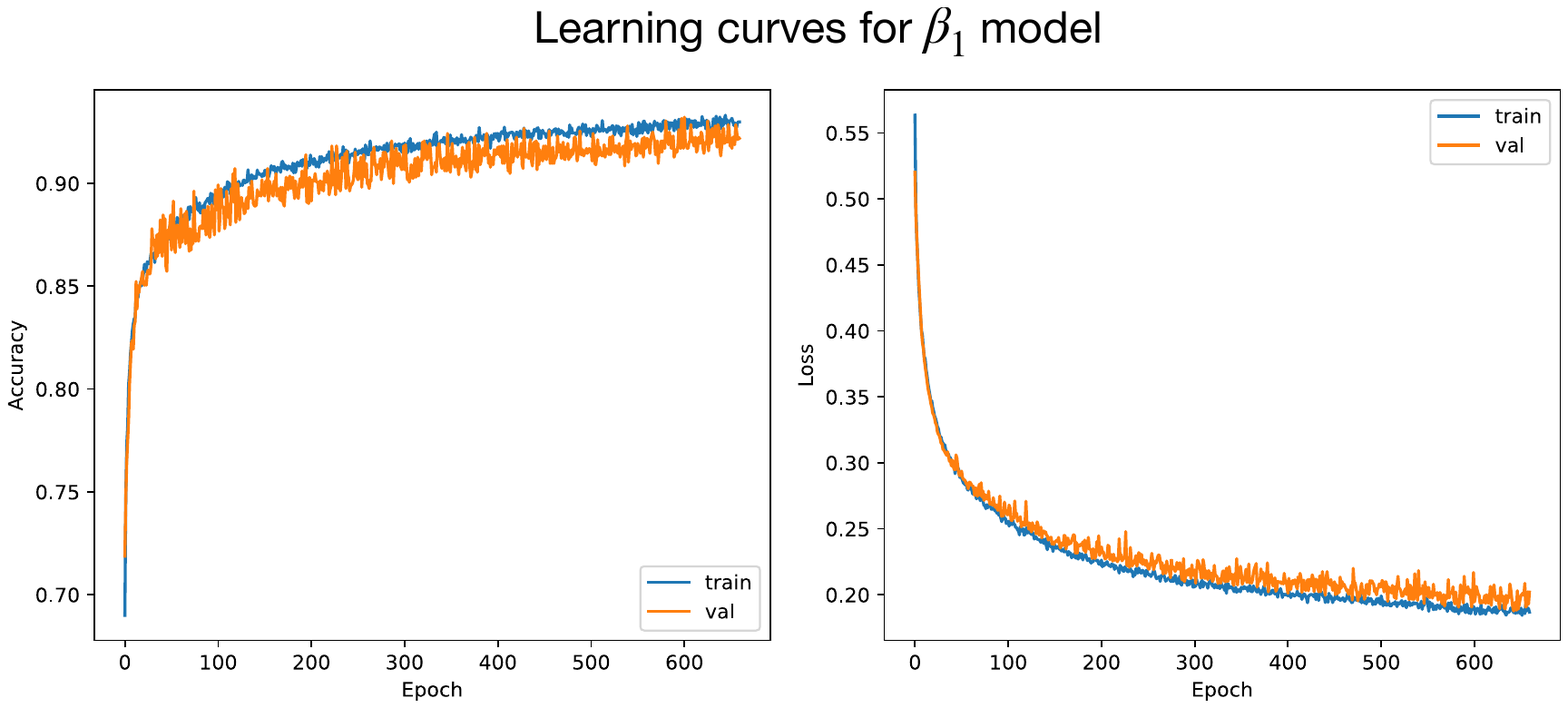}  
\caption{
Left: Accuracy curve for both the training (blue) and validation (orange) datasets for the model where only $\beta_1$ is activated. Right: Its corresponding loss curve.}
\label{fig:learning_curve_beta_1}
\end{figure}

In the case where only the late-time coupling is activated (where $\beta_1$ is non-zero), Fig. \ref{fig:learning_curve_beta_1} shows that our NN performs well, reaching a high accuracy of over $90\%$ and loss of $<20\%$, and is sufficiently trained after just 660 epochs. From the loss curve, we also see that the training and validation losses stabilise and roughly equalise, implying that the model has been able to learn all the features from the training dataset and performs equally well at classifying the unseen validation set.  

\begin{figure*}[ht!]
\begin{subfigure}{.31\linewidth}
  \centering
  \includegraphics[width=1.05\linewidth]{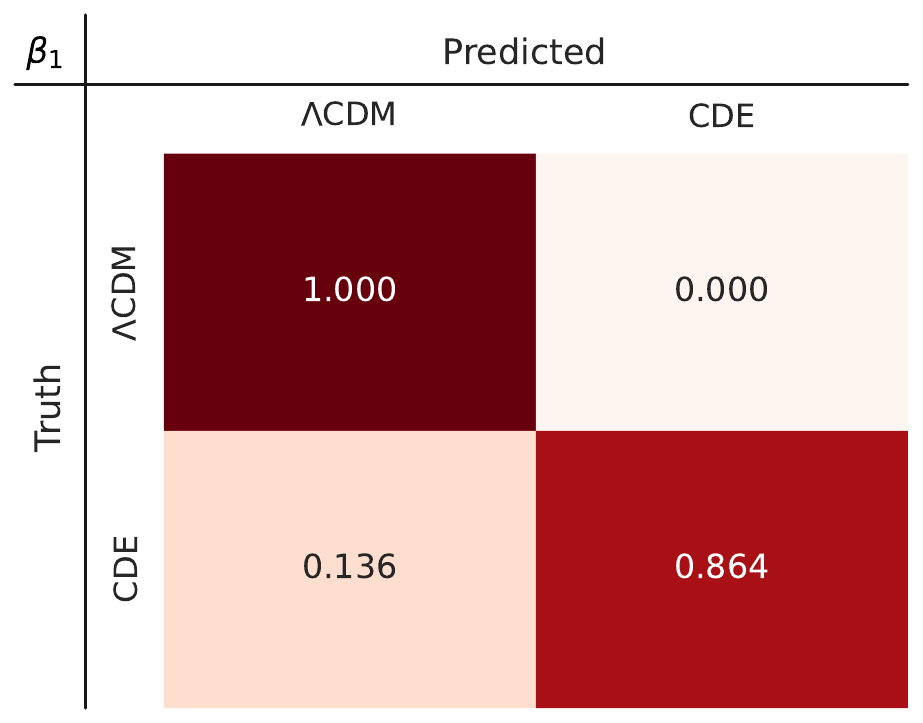}  
  \label{fig:sub-first}
\end{subfigure}
\begin{subfigure}{.31\linewidth}
  \centering
  \includegraphics[width=1.05\linewidth]{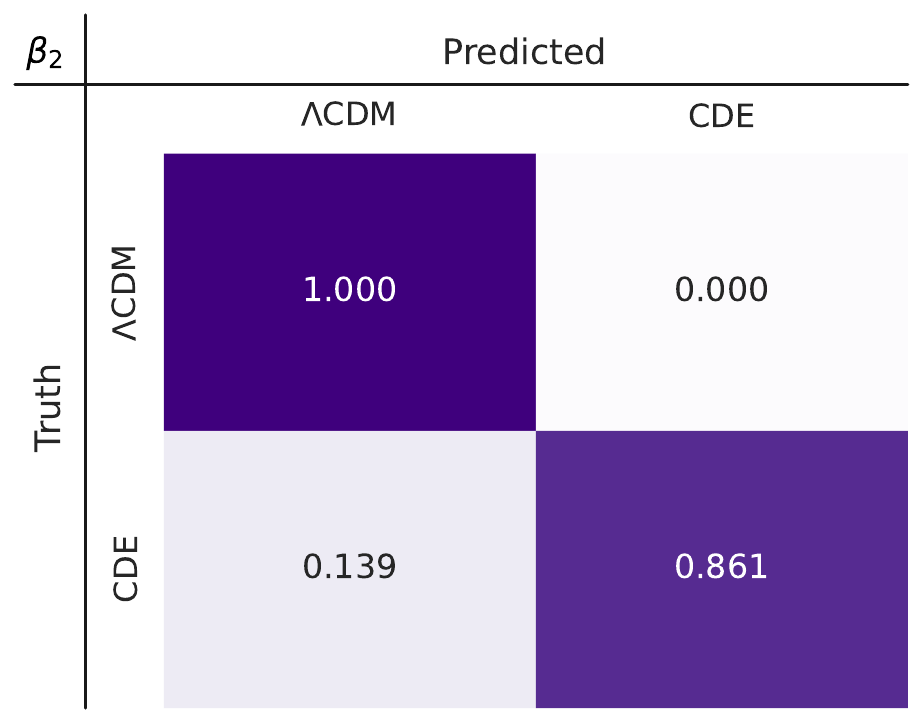}  
  \label{fig:sub-second}
\end{subfigure}
\begin{subfigure}{.31\linewidth}
  \centering
  \includegraphics[width=1.05\linewidth]{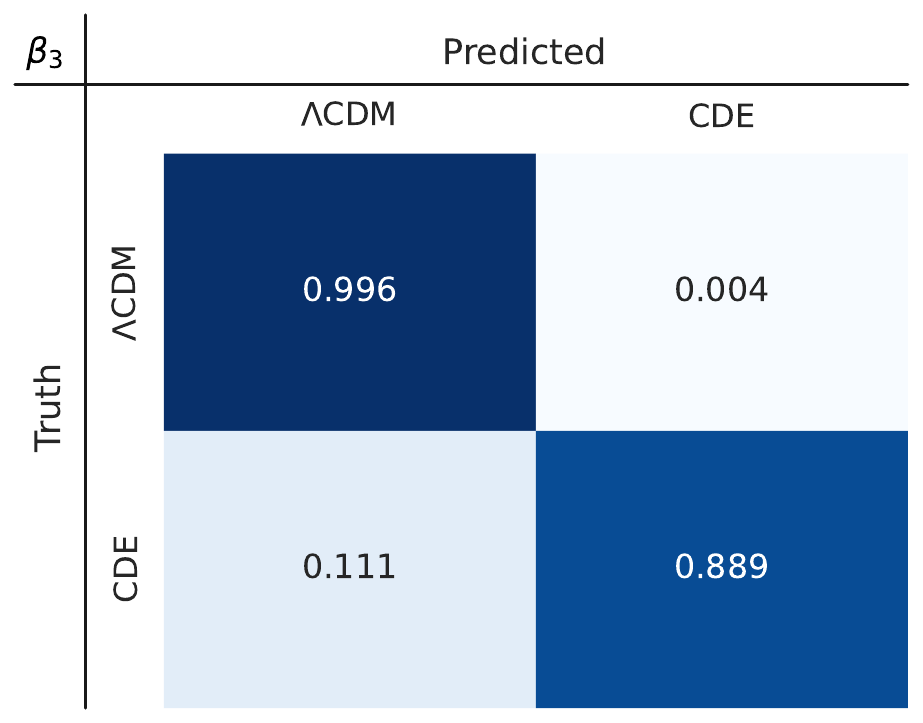}  
  \label{fig:sub-second}
\end{subfigure}
\caption{Classification results in the form of confusion matrices for the cases of switching on coupling $\beta_1$ (left), $\beta_2$ (middle), and $\beta_3$ (right). As a reminder, the tomographic bins for each coupling parameter are $z\,(\beta_1) <100 $, 100 < $z\,(\beta_2) < 1000,$ and $z\,(\beta_3) > 1000$. }
\label{fig:confusion_matrix_single_beta}
\end{figure*}

We also present the classification results of our NN in the form of a confusion matrix in Fig. \ref{fig:confusion_matrix_single_beta}. We see that the NN can accurately predict $100\%$ of the $\Lambda$CDM cases, and also has a high accuracy of $86.4\%$ when classifying CDE cases. 

We see similar performance in the cases of $\beta_2$ and $\beta_3$, where the coupling is activated between redshifts of $ 100 < z < 1000$ and $z>1000,$ respectively. For the $\beta_2$ model, we once again achieve $100\%$ accuracy in identifying $f\sigma_8$ datasets coming from a $\Lambda$CDM model, and $86.1\%$ accuracy for CDE. In the case of $\beta_3$, we see that the accuracy of the NN is $99.6\%$. This might be because activating coupling at early times ($z > 1000$) has the least impact on the increase in $f\sigma_8$, as illustrated in Fig. \ref{fig:fsigma8}. Hence, since the discrepancy between a $\Lambda$CDM dataset and a CDE dataset is marginal, the NN might not have been able to differentiate between the two as accurately as in the cases of $\beta_1$ and $\beta_2$. We present the accuracy and loss curves for these two cases in Appendix \ref{app:c}.

\subsection{Generalisation of $\beta$}
Here we also explore the implementation of an NN architecture capable of performing multi-class classification. For this, we created a dataset from the aforementioned cases of each tomographic bin, with the classes defined as follows: `0' for $\Lambda$CDM, `1' for $\beta_1$ and `2' for $\beta_2$+$\beta_3$, where for the last case we activated these two couplings independently, and combined the generated datasets into one class, where we assumed coupling at high redshifts of $z > 100$. The motivation comes from the fact that we are simulating DESI-like data, where the redshift bins in which this survey operates are between $z \in [0.05,1.65]$. Hence, we do not expect the NN to be able to differentiate between low-redshift data generated with a coupling at $100<z<1000$ and at $z>1000$.

\begin{figure}[t!]
    \centering
    \includegraphics[width=0.55\linewidth]{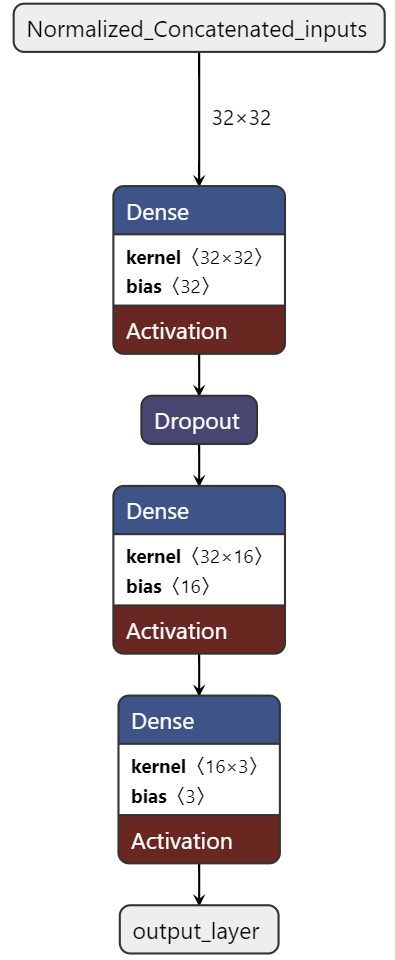}
    \caption{NN architecture implemented for the multi-classification task. The normalisation of features and their concatenation as an input array was performed within the architecture before training. 32$\times$32, as denoted beside the topmost arrow, represents the 32 features with a batch size of 32.}
    \label{fig:NN_architecture_multi}
\end{figure}

We also implemented feature normalisation with a batch size of 32 (see a summary of the architecture in Fig.~\ref{fig:NN_architecture_multi}), while highlighting that we added one more hidden layer with a ReLU activation function and 16 neurons as we found that this improved the multi-class accuracy, as can be seen in Fig.~\ref{fig:mult-beta}. The optimal dropout rate found by \texttt{Optuna} was 0.1 and the number of training epochs obtained by the early stopping callback was 875. Another difference with respect to the previous architectures was that the output layer contained three units for each category and a \texttt{softmax} activation function. We compiled this model using a Nadam optimiser \citep{tato2018improving} as it improved the results in comparison with other optimisers, and a sparse categorical cross-entropy loss function.

\begin{figure}[h!]
  \centering
  \includegraphics[trim={0 5cm 0 2cm},clip,width=\linewidth]{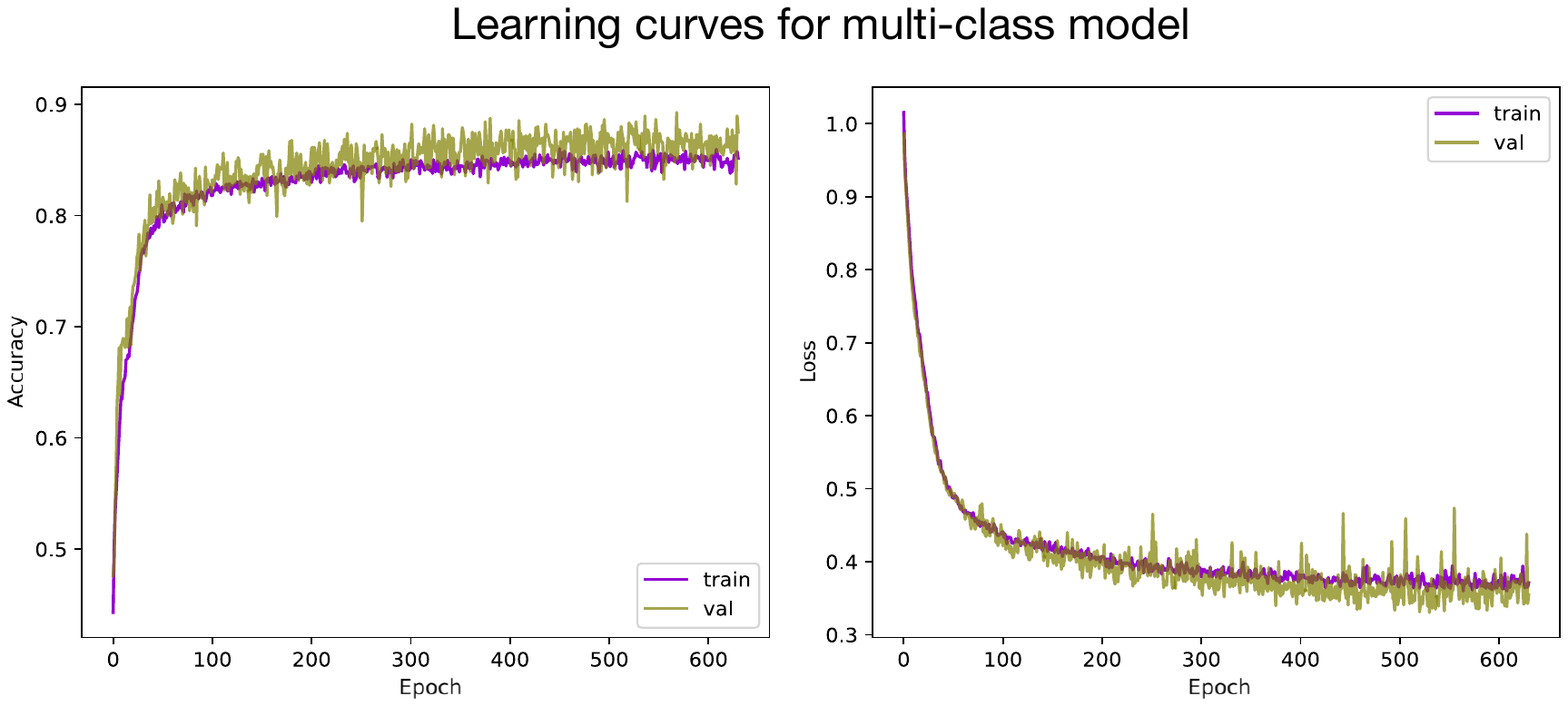} 
\caption{
Left: Accuracy curve for both the training (purple) and validation (green) datasets for the model where the three parameters $\beta_1$, $\beta_2$, and $\beta_3$ are activated. Right: Corresponding loss curve. We considered a three-class classification task, with datasets generated from $\Lambda$CDM, CDE($\beta_1$), and CDE($\beta_2$+$\beta_3$) models.}
\label{fig:learning_curve_multi_beta}
\end{figure}

\begin{figure}
\centering
\includegraphics[width=1.\linewidth]{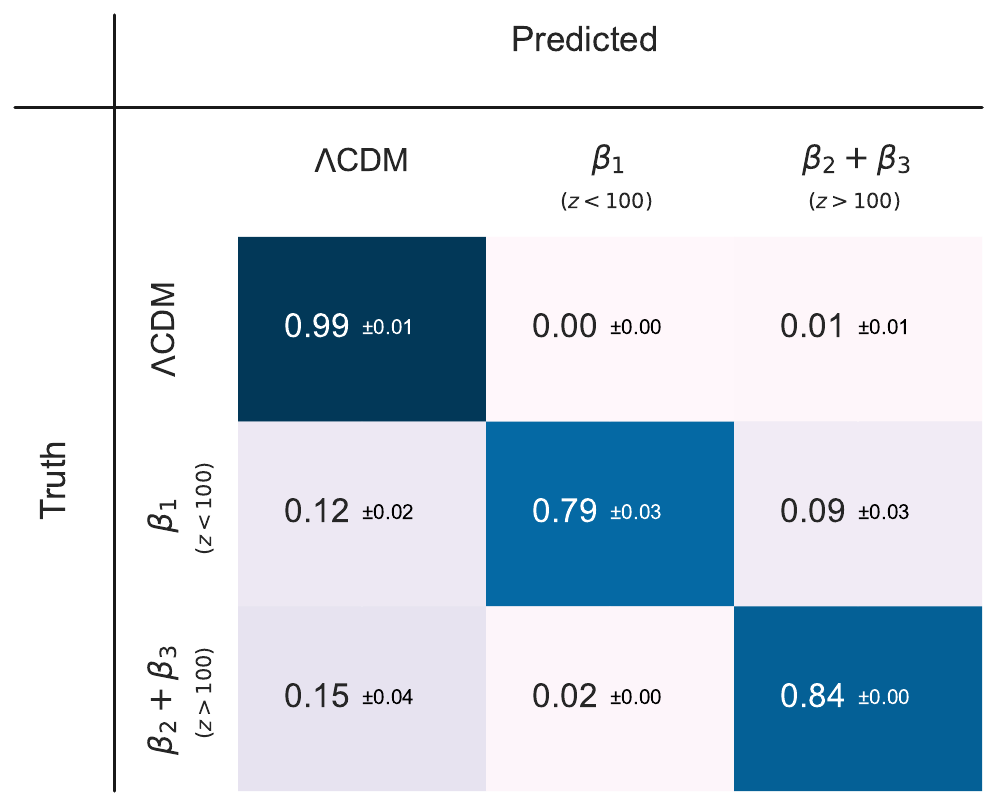}  
\caption{Confusion matrix for the NN multi-classification performance and its errors. In this case, we show the distinction between three classes: $\Lambda$CDM and CDE through the activation of couplings in $\beta_1$ (low-redshift coupling at $z<100$) and in $\beta_2 + \beta_3$ (high-redshift coupling at $z>100$)As a reminder, the tomographic bins for each coupling parameter are $z\,(\beta_1) <100 $, $100 < z\,(\beta_2) < 1000,$ and $z\,(\beta_3) > 1000$.}
\label{fig:mult-beta}
\end{figure}

The learning curves for this multi-class task are illustrated in Fig.~\ref{fig:learning_curve_multi_beta}, where we can see that the accuracy reached by our model is about 85$\%$ and the loss 35$\%$. The results of this multi-class classification task are shown in Fig.~\ref{fig:mult-beta}. We also computed the errors of the predictions performed by the NN, since the impact of randomness in the training was found to be slightly more significant than the previous cases. We can see that the prediction of $\Lambda$CDM data achieves very high accuracy compared to the CDE models. 99$\%$ of the $\Lambda$CDM data samples were correctly classified. The growth data coming from the $\beta_1$ and $\beta_2$+$\beta_3$ activation models are respectively 79$\%$ and 84$\%$ correctly classified, which are lower than in the binary classification scenario. The lowest performance comes from a 15$\%$ of the $\beta_2$ and $\beta_3$ data being mistakenly classified as coming from the $\Lambda$CDM model most likely because, by looking at Fig.~\ref{fig:fsigma8}, the $f\sigma_8$ data coming from both cases are very close to the $\Lambda$CDM case (recalling that it does not show the errors included in the mock data generation). We performed the test again to evaluate whether our architecture could discriminate between four classes instead, by separating class `2' (combination of $\beta_2$ and $\beta_3$ data) into `2' ($\beta_2$ data only) and `3' ($\beta_3$ data only); however, the NN poorly differentiates the data. This is to be expected since we do not anticipate being able to probe dynamics at high redshifts with DESI-like data, which fall within the redshift range $z=[0.05,1.65]$. We argue that our architecture being able to differentiate $f\sigma_8$ data coming from the activation of low- and high-redshift tomographic couplings is a substantial result in itself.

\section{Conclusions \label{sect:conclusion}}
In this work, we explored whether NNs can differentiate between two different cosmological models, namely the standard $\Lambda$CDM model and the tomographic CDE model, where a coupling exists between DM and the DE scalar field.
To do so, we used mock growth-rate ($f\sigma_8$) data based on realistic DESI-like survey specifications created using a Fisher matrix approach. These realisations of the mock data were then used to train, validate, and test the NN. In the case of the CDE model, we also considered a three-bin tomographic coupling $\beta$ that depends on the redshift $z$, and is assumed to be constant (and independent) in the three bins in order to capture a possible evolution of the coupling.

After creating the mocks, we explored different NN architectures (see Fig.~\ref{fig:NN_architecture} for a visual summary and  Sect.~\ref{sec:NN_arc}). We find that when treating each parameter of the coupling, separately by assuming only one of the $\beta$s is free while fixing the rest to zero, the NN can achieve nearly perfect classification for the $\Lambda$CDM model and approximately $86-89\%$ accuracy for the CDE for each of the $\beta$ models (see the confusion matrices in Fig.~\ref{fig:confusion_matrix_single_beta}). Using the NN optimisation package \texttt{Optuna}, we ensured that our NN was well optimised, with the best number of layers and hyperparameters to prevent overfitting (see Appendix \ref{app:b} for more details).

We also considered a multi-class classification scenario, we conducted a three-class classification task, investigating whether the NN can simultaneously differentiate between data originating from $\Lambda$CDM, CDE with low-redshift coupling, and CDE with high-redshift coupling. We did this by combining the data coming from the activation of $\beta_2\,(100<z<1000)$ and $\beta_3\,(z>1000)$, since our DESI-like setup only contains data at low redshifts and the $\beta_2$ and $\beta_3$ models involve coupling at high redshifts where degeneracies with other cosmological parameters may limit the predictive accuracy of the NN. This proved more demanding for the architecture in the previous case, necessitating the addition of an extra hidden layer with a ReLU activation function and 16 neurons to improve the multi-class task (see Fig.~\ref{fig:NN_architecture_multi} for more details). We find that the classification accuracy for a $\Lambda$CDM model is nearly perfect (around $99\%$), while we achieved a performance of around $79\%$ for $\beta_1$ and $84\%$ for $\beta_2$+$\beta_3$.

To verify the robustness of our analysis, we also performed several tests on our NN architecture as described in Appendices \ref{app:a}, \ref{app:b}, \ref{app:c}, and \ref{app:d}, where we find the optimal number of training datasets, use \texttt{Optuna} to optimise the various hyperparameters of the NN, present the accuracy and loss curves for the $\beta_2$-$\beta_3$ models, and use the Akaike information criterion as a complementary classification analysis, respectively.

Finally, an interesting question is whether our NN could in principle distinguish our CDE model from other variations, for example those presented in \cite{Gumjudpai:2005ry}, \cite{Gavela:2010tm}, \cite{Salvatelli:2013wra}, \cite{Amendola:2006dg}, and \cite{Pourtsidou:2013nha}.
While a concrete answer would require full numerical simulations and a detailed comparison of the various models, which is something beyond the scope of this work, we can at least speculate on the outcome based on the behaviour of the models. For example, comparing our Eq.~\eqref{eq:consrhoscf_cde} with Eq.~(4) in \cite{Salvatelli:2013wra}, we see that the right-hand sides (RHSs) of the two conservation equations are markedly different. In the first case, the RHS depends on a function of redshift, $\beta(z)$, the DM density, and the derivative of $\phi$, while in the second case, it depends on the conformal Hubble parameter and the dark energy density. Therefore, the two functions have different time dependences and thus affect early- and late-time physics differently.

For example, this can be seen by comparing Fig. 3 of \cite{Goh:2023mau} and Fig. 1 of \cite{Salvatelli:2013wra}, where the effect on the cosmic microwave background peaks, which are normally affected by the DM density, is markedly different by more than a few percent, due to the different dependence on the evolution of the RHS of the continuity equations. As a result, we expect a similar effect on the evolution of the growth rate in both models. Thus, we can expect our NN to be able to discriminate between the two models, although we leave a more detailed analysis and comparison for the future.

In summary, our work highlights the advantages of employing deep learning techniques, in this case, an NN pipeline, to analyse spectroscopic growth-rate ($f\sigma_8$) data from current Stage IV surveys like DESI and potentially other LSS experiments, especially when testing models with DM and DE couplings, which can have degeneracies at high redshifts and can be difficult to disentangle from the cosmological constant model. We find that our pipeline can confidently (with $>86\%$ accuracy) detect non-zero values of the $\beta$ coupling at some redshift range and with $100\%$ confidence detect the $\Lambda$CDM model. It can thus be a useful tool for analysing these data and maximising the potential of current surveys to probe for deviations from general relativity.


\begin{acknowledgements}
IO thanks ESTEC/ESA for the hospitality during the execution of this project, and for support from the ESA Archival Research Visitor Programme. LG thanks ESTEC/ESA and the Instituto Física Teórica for their hospitality and financial support for her visits in April and June 2024. IO and SN acknowledge support from the research project PID2021-123012NB-C43 and the Spanish Research Agency (Agencia Estatal de Investigaci\'on) through the Grant IFT Centro de Excelencia Severo Ochoa No CEX2020-001007-S, funded by MCIN/AEI/10.13039/501100011033. IO is also supported by the fellowship LCF/BQ/DI22/11940033 from ``la Caixa” Foundation (ID 100010434) and by a Graduate Fellowship at Residencia de Estudiantes supported by Madrid City Council (Spain), 2022-2023. SN and IO acknowledge the use of the Finis Terrae III Supercomputer which was financed by the Ministry of Science and Innovation, Xunta de Galicia and ERDF (European Regional Development Fund), also the IFT UAM/CSIC cluster Hydra and the \textit{San Calisto} supercomputer, courtesy of M.~Martinelli. This work was made possible by utilising the CANDIDE cluster at the Institut d’Astrophysique de Paris, which was funded through grants from the PNCG, CNES, DIM-ACAV, and the Cosmic Dawn Center and maintained by S. Rouberol.
\end{acknowledgements}

\section*{Data availability}
The NN and the relevant scripts used to conduct the analysis are publicly available as a GitHub repository at the link: \url{https://github.com/IndiraOcampo/Growth_LSS_model_selection_CDE.git}. 

\bibliographystyle{aa}
\bibliography{ref}


\begin{appendix}
\section{Number of training samples} \label{app:a}
\begin{figure}[h!]
    \centering
    \includegraphics[width=\linewidth]{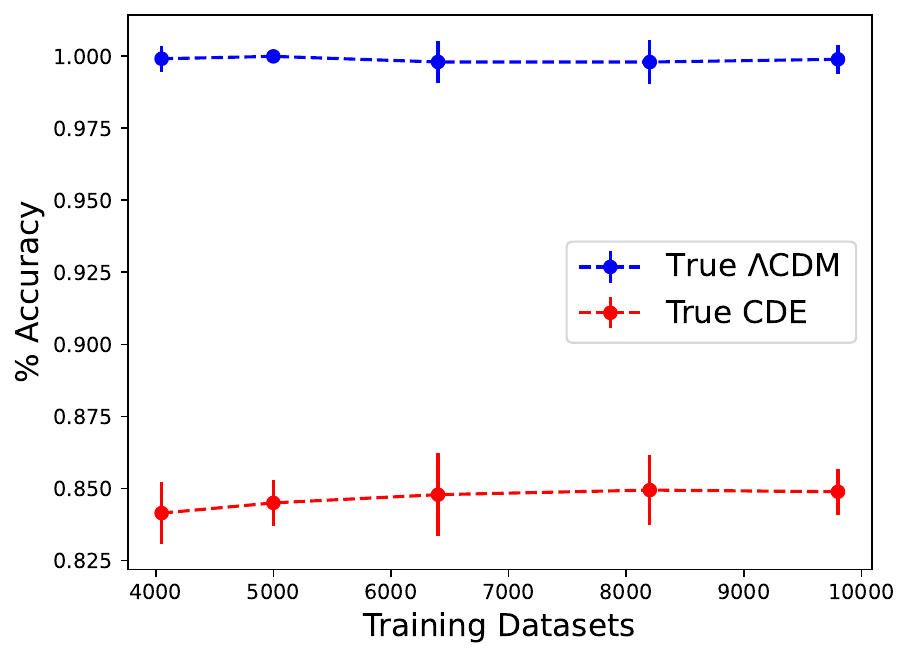}
    \caption{Percent accuracy of the NN against the number of training datasets used. The circular points mark the average values out of 50 runs, with the error bars denoting the 1$\sigma$ standard deviation.  }
    \label{fig:trianing_samples_perf}
\end{figure}

Here we show several tests of the effect of increasing the number of training datasets to [4050, 5000, 6400, 8200, 9800] where half of the dataset are samples from $\Lambda$CDM, and the other half is generated from a CDE model. From Fig. \ref{fig:trianing_samples_perf} we see that the NN is able to accurately distinguish almost 100\% of the $\Lambda$CDM cases regardless of the number of training datasets. In the CDE case, results improve with an increasing number of datasets up to an optimum value of about 8000, after which performance plateaus, indicating that the network does not learn any more new information with the increase in the number of data points.

\section{Optimisation with \texttt{Optuna}\label{app:b}}
In Sect. \ref{sec:results} we use \texttt{Optuna} to optimise our NN for each CDE model ($\beta_1, \beta_2$ and $\beta_3$). Here, we assess the effectiveness of implementing this additional optimisation step by also testing the NN when it was not optimised, namely using an arbitrarily chosen fiducial architecture of one hidden layer, 32 nodes, a dropout fraction of 0.5 and fixing the number of training epochs at 2000. We present our results in Table \ref{tab:opt}.
\begin{table}[h]
\renewcommand{\arraystretch}{1.2}
  \centering
    \caption{Percentage accuracy of the NN classification scheme when no optimisation is implemented (middle two columns) as compared to when the NN is optimised with \texttt{Optuna} (rightmost two columns) for all three models studied.}
  \begin{tabular}{| c |  c | c | c | c | }
    \hline \hline
      \multirow{2}{*}{Model} & \multicolumn{2}{c|}{No optimisation} & \multicolumn{2}{c|}{Optimisation with \texttt{Optuna}} \\
      \cline{2-5}
                             &$\Lambda$CDM &CDE& $\Lambda$CDM &CDE\\
     \hline
    $\beta_1$  &1.000& 0.843 & 1.000 & 0.864\\
    \hline
    $\beta_2$  &1.000& 0.836 & 1.000& 0.861\\
    \hline
    $\beta_3$  &1.000& 0.835& 0.996 & 0.889\\
    \hline \hline
  \end{tabular}
  \label{tab:opt}
\end{table}

We see that optimisation improves our results in all three cases. This might be because this is a relatively straightforward classification problem requiring a simple network architecture, and the optimised set of hyperparameters as reported in Table \ref{tab:hyperparams} did not deviate much from the non-optimised vanilla setup, which was already adequate for this problem. We expect the optimisation to prove more impactful for complex problems requiring multiple hidden layers. Nevertheless, we have demonstrated the robustness of our NN architecture. 

\section{Accuracy and loss performance}\label{app:c}
Here we present the accuracy and loss curves of the training and validation sets, for the models where we turned on coupling at redshifts $100<z<1000$ (activating only $\beta_2$) and $z>1000$ (activating only $\beta_3$). We see similar results in all three models, with accuracy reaching beyond 90\%.

\begin{figure}[h!]
  \centering
  \includegraphics[trim={0 5cm 0 2cm},clip,width=\linewidth]{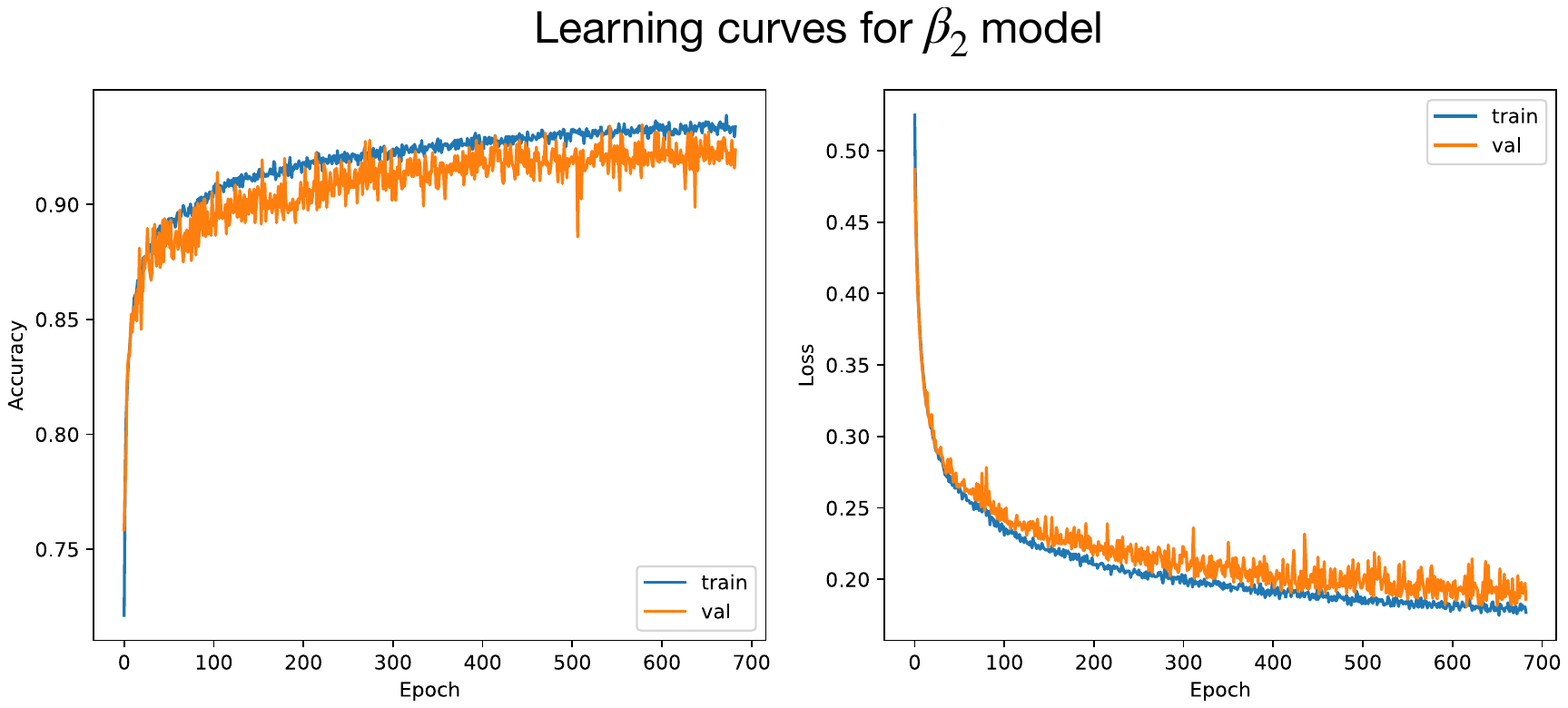} 
\caption{
Left: Accuracy curve for both the training (blue) and validation (orange) datasets for the model where only $\beta_2$ is activated. Right: Its corresponding loss curve.}
\label{fig:learning_curve_beta_2}
\end{figure}

\begin{figure}[h!]
  \centering
  \includegraphics[trim={0 5cm 0 2cm},clip,width=\linewidth]{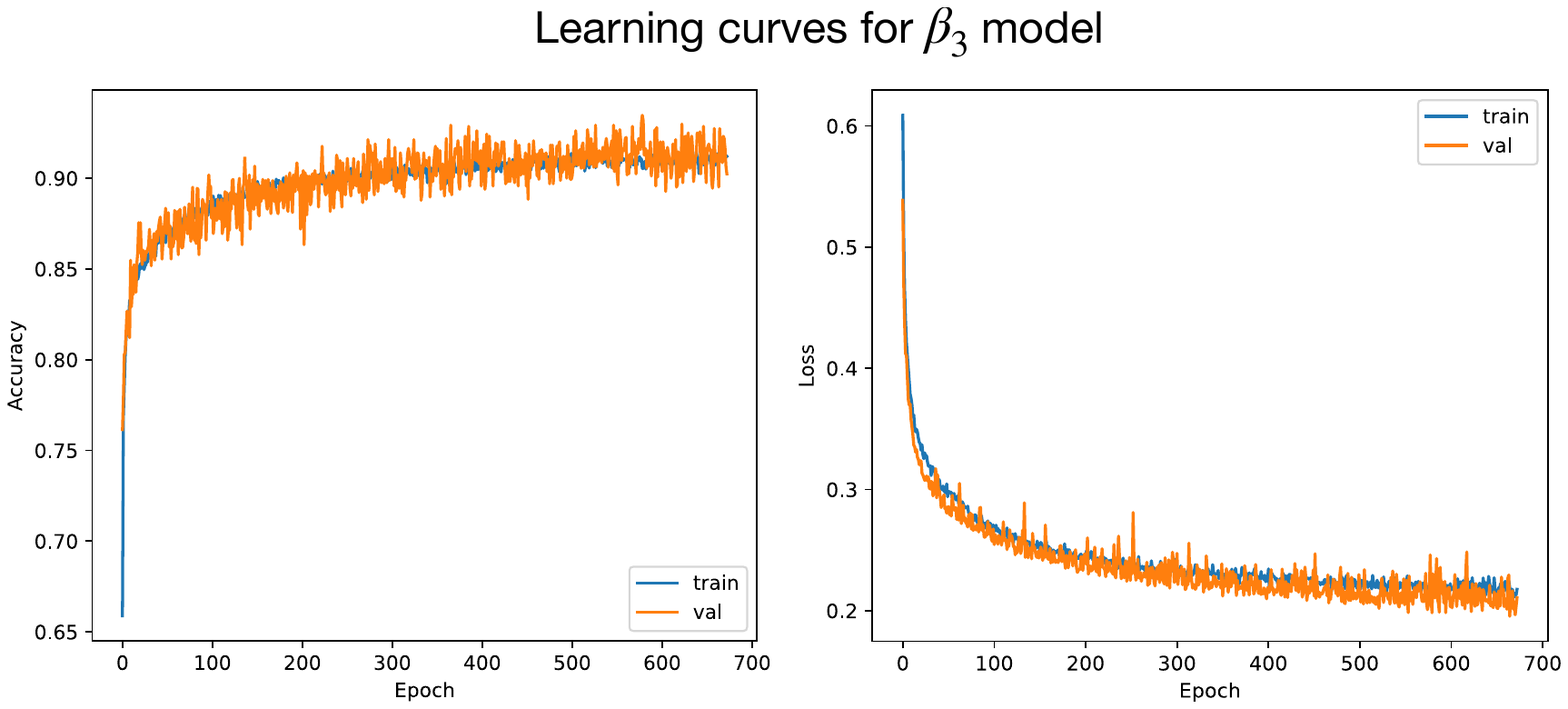}  
\caption{Same as Fig. \ref{fig:learning_curve_beta_2} but for the model where only  $\beta_3$ is activated. }
\label{fig:learning_curve_beta_3}
\end{figure}

\section{Complementary Bayesian analysis}\label{app:d}

Here we briefly present a complementary Bayesian analysis in the case of the CDE model where $\beta_1$ is free to vary as in Sect.~\ref{sec:results_beta}, using the corrected Akaike information criterion (AICc; see \citealt{Akaike1974}). Specifically, under the assumption of Gaussian errors, the estimator is
described via
\begin{eqnarray}
{\rm AICc} = -2 \ln {\cal L}_{\rm max}+2k_p+\frac{2k_p(k_p+1)}{N_{\rm dat}-k_p-1} \label{eq:AIC}\;,
\end{eqnarray}
where $N_{\rm dat}$ and $k_p$ denote the total number of data points and the number of free parameters (see also \citealt{Liddle:2007fy}). In our case, we have 16 data points for each of the $f\sigma_8$ realisations and the number of parameters we varied is $\Omega_c h^2$ for $\Lambda$CDM and [$\Omega_m, \beta_1$] in the CDE model; thus $k_p=1$ and $k_p=2,$ respectively, for the two models.

To compare the two cosmological models, namely the CDE and the $\Lambda$CDM, we then introduced the quantity $\Delta {\rm  AICc}\equiv{\rm AICc}_{\rm model}-{\rm AICc}_{\rm min}$, which is the relative difference of the AICc estimators and can be interpreted as follows \citep{burnham2002model}: if $\Delta {\rm AICc} \le 2$ then the two models are statistically consistent, if $2<\Delta {\rm AICc} <4$ there is weak evidence in favour of the model with the smallest AICc, while if $4<\Delta {\rm AICc} <7$ then there is definite evidence against the model with higher value of ${\rm AICc}$, while finally, if $\Delta {\rm AICc} \ge 10$ then this suggests strong evidence against the model with the higher AICc.

Then we calculated the $\Delta {\rm  AICc}$ values for all the realisations used for the testing of the NN architecture, corresponding to different values of the cosmological parameters in the grid, in the case of a varying $\beta_1$. Doing so we find that the difference in the AICc estimator between CDE and $\Lambda$CDM overall is $\Delta {\rm  AICc}\simeq 2.59\pm0.16$, signifying weak evidence in favour of $\Lambda$CDM over the CDE, even if some of the mocks were created assuming the latter model.

\end{appendix}


\end{document}